\def\etal{{\rm et al. }}

\def\kms{{\rm km s^{-1}}}
\newcommand\aap{{\em A}\&{\em A}}

\newcommand\aj{{\em AJ}}
\newcommand\apj{{\em ApJ}}

\newcommand\apjs{{\em ApJS}}
\newcommand\araa{{\em ARA}\&{\em A}}

\newcommand\mn{{\em MNRAS}}

\newcommand\nat{{\em Nature}}
\newcommand\nature{{\em Nature}}
\newcommand\pasp{{\em PASP}}

\documentclass[structabstract]{aa}  
\usepackage{graphicx}
\usepackage[sort]{natbib}
\bibpunct{(}{)}{;}{a}{}{,}
\begin{document}

   \titlerunning{Effect of bars in AGN host galaxies and black hole activity}
   \authorrunning{Alonso et al.}

   \title{Effect of bars in AGN host galaxies and black hole activity}

   \subtitle{}

   \author{Sol Alonso\inst{1,2},
          Georgina Coldwell\inst{1,2}
   		  \and
          Diego G. Lambas \inst{1,3}
          }

   \institute{Consejo Nacional de Investigaciones Cient\'{\i}ficas y
T\'ecnicas, Argentina\\
              \email{salonso@icate-conicet.gob.ar}
         \and
             ICATE, CONICET-UNSJ, CC 49, 5400, San Juan, Argentina\\
            \and
IATE, CONICET, OAC, Universidad Nacional de C\'ordoba, Laprida 854,
X5000BGR, C\'ordoba, Argentina
             }
             
   \date{Received xxx; accepted xxx}

  \abstract
   {}
   { With the aim of assessing the effects of  bars on active galactic nuclei (AGN), we present an analysis of  host characteristics and nuclear activity of AGN galaxies with and without bars.}
   { We select AGN galaxies from the Sloan Digital Sky Survey Data Release 7 (SDSS-DR7), using the available emission-line fluxes. By visual inspection of SDSS images 
we classified the hosts of face-on AGN spiral galaxies brighter than $g-$mag$<$16.5 into barred or unbarred.
With the purpose  of providing an appropriate quantification of the effects of bars, 
we also constructed a suitable control sample of unbarred AGN galaxies 
with similar redshift, magnitude, morphology, bulge sizes and local environment
distributions.}
   {We find that the bar fraction, with respect to the full sample of 
spiral face-on AGN host galaxies, is 28.5$\%$,
in  good agreement with previous works. 
Barred AGN host galaxies show an excess of young stellar populations (as derived with the 
$D_n(4000)$ spectral index), dominated by red $u-r$ and $g-r$ colors, with respect to the control 
sample, suggesting that bars produce an important effect on galaxy properties of AGN hosts.
Regarding the nuclear activity distribution, we find that barred AGN galaxies show a shift toward higher $Lum[OIII]$ values with respect to AGN without bars.
In addition, we also find that this trend is more significant in less massive, younger stellar population and bluer AGN host galaxies.
We found that the fraction of powerful AGN increase 
towards more massive hosts with bluer colors and younger stellar populations residing in denser environments. However, barred host AGN systematically show a higher fraction of powerful 
active nuclei galaxies with respect to the control sample.
We also explored the accretion rate  onto the central black holes finding 
that barred AGN show an excess of objects with high accretion rate values with respect to unbarred AGN galaxies. 
}
   {}


   \maketitle
%

\section{Introduction}

It is widely accepted that the mechanism responsible for Active Galactic Nuclei (AGN)
is based on the accretion of matter by massive black holes at the centres
of galaxies \citep{LB69,Rees84}.
Several theories have been proposed to explain the black hole feeding mechanisms
needed to transport mass to the most inner region of the central engine. The major galaxy interaction
\citep{alonso07},  minor mergers \citep{roos81,roos85,HM95}, tidal disruption
produced by galaxy harassment \citep{LKM98}, gas dynamic processes involving the presence of
multiple black holes \citep{BBR80} and gravitational instabilities in the disk of barred galaxies
\citep{SBF90} can cause gas transfer to the central regions of galaxies.
One must also consider that it is plausible that at any given time or for any given AGN type, more
than one mechanism plays a role.

In particular, the presence of bars plays an important role in the evolution, star-formation and
nuclear activity of galaxies. The relation between star-forming galaxies and bars has
been the subject of several works \citep{balza83,debe87} finding that
in some cases bars seem to be an appropriate way to produce radial gas flow favoring an increasing
star formation rate in the galaxies \citep{martin95} which could be even more effective
than galaxy interactions  \citep{Elli11}.
Considering the known connection between star-formation and nuclear activity of
galaxies \citep{sprin05,dimat05} it could be expected that bars also increase the AGN activity.

Furthermore, several studies suggest that the gas clouds within galaxies suffer
shocks by interaction with the edges of the bars producing 
angular momentum losses and allowing a flow of gas within the innermost regions
of galaxies \citep{SBF90}. Also, a system of bars ("bars within bars")
of different lengths and properties has been proposed \citep{SBF90}
and observed \citep{Emse01} to have an association with the AGN activity.
\cite{erw04} found fifty double-barred galaxies with inner or secondary bar is embedded within a large-scale, outer or primary bar, of the
catalog of 67 barred galaxies which contain elliptical stellar structures inside their bars.
In addition, the presence of bars can explain the formation of bulges
which are also directly related to the probability of occurrence of AGN \citep{wyse04}.
\cite{coel11} analyzed barred and unbarred galaxy samples with similar stellar mass distributions, finding similar age-metallicity diagrams in both samples; however, in the distribution of bulge ages in barred galaxies showed an excess of younger populations, with respect to bulges in unbarred ones.
They also found, for low-mass bulges, twice as much AGN in barred galaxies, as compared to unbarred galaxies.

However,  although there is observational evidence that bars cause central
concentrations of gas, compared to non-barred galaxies, there is no clear agreement
whether the large-scale bars are efficient for the transport of
material directly to the regions closest to the black hole \citep{knapen05}.
\cite{HFS97} found that the existence of bars in a sample of 300 spiral galaxies 
has no significant effect on the AGN power.
More recently, \cite{oh12}  found a signal of bar effect 
in the AGN activity mainly present in blue galaxies with low
black hole mass. This AGN sample was selected with \cite{kauff03} criteria from 
a volume limited sample of SDSS-DR7 late-type galaxies ($a/b > 0.6$) with good image qualities.
On the other hand, \cite{lee12} used a sample of AGN, selected with  \cite{kew01} classification,
also from SDSS-DR7 volume limited sample of SDSS-DR7 late-type galaxies ($a/b > 0.7$) 
finding that AGN power is not enhanced by bars.
In this analysis, \cite{lee12} do not include the Composite galaxies within the sample of AGN 
host and these galaxies are, in general, younger and less massive \citep{kew06} than the AGN 
sample defined by the \cite{kew01} criteria. So, the different AGN selection criteria could 
affect the results related to the effect of enhanced AGN activity  by bars.

This discrepancy in the results could be due mainly to: i) a biased selection of the control 
sample, used for the comparison of the obtained results. In this sense \cite{perez09} 
suggest that a control sample for interacting galaxies 
should be selected matching, at least, redshift, morphology, stellar masses and local densities. 
This criteria is, also, applicable to the building of control samples of barred galaxies due 
to its having different properties than unbarred galaxies.
ii) the short time of life of bars, 1-2 Gyr \citep{bour05} which represents 
a brief period in the life of an AGN. Moreover, studies using numerical 
simulations \citep{bourco02} found that bars within galaxies can be
destroyed and re-formated a couple of time during the galaxy life. 
Moreover, the 3D N-body simulations \citep{altha05} showed that the central mass concentration necessary to destroy the bar is, at least, several per cent of the mass of the disc;  suggesting that, supermassive black holes are not likely to destroy pre-existing bars. 
So, in this context bars 
formed at different times, with different conditions,  could be present in the sample of barred galaxies.

In this work we made a detailed study of the properties of barred host galaxies and
the effects that bars produce in the AGN power is a sample of spiral galaxies comparing the results
with a carefully selected sample of unbarred AGN. In section 2 we described the type II AGN 
selection method, the bar classification and the control sample selection criteria. In 
section 3 we analyse the dependence of the AGN host galaxy properties with the bar presence. 
The influence of bars in the black hole activity is detailed in Section 4 and finally the 
conclusions are summarized in Section 5. The adopted cosmology through this paper is:  
$\Omega = 0.3$, $\Omega_{\lambda} = 0.7$  and $H_0 = 100~ \kms \rm Mpc$.


\section{The AGN SDSS-DR7 Catalog} 

The Sloan Digital Sky Survey (SDSS, \cite{york00}) is the largest galaxy 
survey at the present.
It uses a 2.5m telescope \citep{gunn06} to obtain photometric and 
spectroscopy data that will cover approximately one-quarter of 
the celestial sphere and collect spectra of more than one million objects.
The seven data release imaging (SDSS-DR7, \cite{abaz09}) comprises 11663 
square degrees of sky imaged in five wave-bands ($u$, $g$, $r$, $i$ and $z$) containing
photometric parameters of 357 million objects. 
SDSS-DR7 therefore represents the final data 
set that will be released with the original targeting and galaxy selection 
\citep{Eisens01,strauss02}.
The main galaxy sample, which contains $\sim 700000$ galaxies with 
measured spectra and photometry, is essentially a magnitude limited spectroscopic sample 
(Petrosian magnitude) \textit{$r_{lim}$}$ < 17.77$, most of galaxies span a
redshift range $0 < z < 0.25$ with a median redshift of 0.1 \citep{strauss02}.

Several physical galaxy properties have been derived and published for the SDSS-DR7 
galaxies: gas-phase metallicities, stellar masses, 
indicators of recent major star-bursts, current total and specific 
star-formation rates, emission-line fluxes, S\'ersic index, etc. 
\citep{brinch04, tremonti04, blanton05}.
These data were obtained from the MPA/JHU\footnote{http://www.mpa-garching.mpg.de/SDSS/DR7/} and 
the NYU\footnote{http://sdss.physics.nyu.edu/vagc/} value added catalogs. 

For the AGN selection we used the publicly available emission-line fluxes. The method
for emission-line measurement is detailed in \cite{tremonti04}.
Additionally, we corrected the emission-line fluxes for optical reddening using
the Balmer decrement and the \cite{calzetti00} obscuration curve.
We assumed an $R_V=A_V/E(B-V)=3.1$ and an intrinsic $H\alpha/H\beta=3.1$.
Since the true uncertainties in the emission-line measurements were underestimated
the signal-to-noise ($S/N$) of every line was calculated with the emission-line flux 
errors adjusted according to the uncertainties suggested by the MPA/JHU 
catalog \footnote{http://www.mpa-garching.mpg.de/SDSS/DR7/raw$\_$data.html}.

The AGN galaxy sample was selected using a standard diagnostic diagram proposed by \cite{BPT81} 
(hereafter BPT). This diagram allows for the separation of type 2 AGN from normal star-forming 
galaxies using emission-line ratios and depending on their position in the diagram.
Furthermore, we used only galaxies with signal-to-noise ratio S/N $> 2$ for all the lines
intervening in the diagnostic diagram used to discriminate AGN from HII galaxies.
This S/N cut was selected taking into account that the adjusted uncertainties almost duplicated
the original errors.
So, taking into account the relation between spectral lines, $\rm [OIII]\lambda 5007$, $\rm H\beta$,
$\rm [NII]\lambda 6583$ and $\rm H\alpha$, within the BPT diagram we follow the \cite{kauff03} 
criterion to select type 2 AGN as those with:   

\begin{equation}
\log_{10}([OIII]/\rm H\beta) > 0.61/(\log_{10}(\rm [NII/H\alpha])-0.05)+1.3,
\end{equation}

\subsection{Selection of barred AGN galaxies}

With the aim of obtaining an AGN sample with barred galaxy hosts, we firstly cross-correlated 
the SDSS-DR7 AGN 
galaxies with the spiral objects obtained from the Galaxy Zoo catalog \citep{lintott11}, 
which comprises a morphological classification of nearly 900,000 galaxies drawn from the SDSS.
Then, we restricted the AGN spiral edge-on galaxy sample in redshift ($z<0.1$) and we imposed a 
magnitude cut such as the extinction corrected SDSS $g-$mag is brighter than 16.5. We have 
selected galaxies with axial ratio $b/a>0.4$ since the efficiency of the visual classification 
decreases with galaxy inclination.
With these restrictions, our AGN sample in the SDSS-DR7 comprises 6772 galaxies. 

By using the photometric SDSS-DR7, we classified the host galaxies of the AGN spiral 
galaxy sample morphologically taking into account the naked-eye detection of barred features, finding 1927 AGN hosted in barred galaxies.
This represents a fraction of 28.5$\%$ with respect to the full sample of 6772 AGN in spiral host
galaxies, and this value agrees with the bar fraction found by visual inspection 
of optical galaxy samples in previous works. 
207 objects which were not completely confiable with the bar classification were not included.
The details of the classification are listed in Table 1. 
We have checked for possible dependencies of our classification on redshift by computing the fraction of barred hosts in different redshift intervals finding a nearly constant value of $\sim 30\% $. 
In particular, there are 244 galaxies within z$<$0.02 out of which 74 (30$\%$) have bars. 
Thus, the presence of small bars, visible in this closer subsample, is not expected to be a 
significant fraction in our sample of unbarred objects.

Several studies, carried out by means of visual inspection of different optical galaxy samples 
(e.g. the RC3 and UGC, \citet{deVau91,nilson73,marinova09}) 
found a bar fraction of 25-30$\%$. 
Moreover, \citet{master10} computed the mean bar fraction, finding 29.4$\%$ 
of barred galaxies from a sample of 13665 Galaxy Zoo disk objects.
More recently, \citet{oh12} detected a bar fraction of 29.5$\%$ (715 AGN barred galaxies), in 
the sample of AGN galaxies obtained from SDSS-DR7, in the redshift range $0.01 < z < 0.05$.
However, others studies using near-infrared images from 2MASS \citep{menendez07} and SDSS image decompositions of $g$, $r$ and $i$ bands \citep{gado09} found a larger fraction of bars, 59\% and 42\%, respectively.

\begin{table}
\center
\caption{Results of classification.}
\begin{tabular}{|c c c| }
\hline
Classification & Number of galaxies &  Percentages  \\
\hline
\hline
Barred             &  1927    &   28.5$\%$     \\
Unclear barred     &  207     &    3.0$\%$     \\
Unbarred           &  4638    &   68.5$\%$     \\
\hline
Total              & 6772     &   100.0$\%$     \\
\hline
\end{tabular}
{\small}
\end{table}

To corroborate the accuracy of our selection criteria, we cross-correlated our sample with barred AGN galaxies taken from the catalog of \cite{nair10a}.
They constructed a catalog of detailed visual classification for 14034
galaxies in the Sloan Digital Sky Survey Data Release 4 (SDSS-DR4).
The sample includes galaxies in the redshift range $0.01 < z < 0.1$, with $g-$mag $<$ 16. They detected 
a total of 454 barred AGN galaxies sub-classified, depending on the relatively light contribution of the bar in:
strong, intermediate and weak. 
We found 399 (87.9$\%$) of a common barred AGN galaxies in the two samples. 
However, the \cite{nair10a} catalog classified 307 strong and 
intermediate barred AGN 
galaxies, from the total sample, from which we found 297 common galaxies, representing the 96.5$\%$ 
of overlap. 
Considering that we have matched catalogs derived
through visual classification, and that this kind of classification could be subjective, the level of agreement 
is very high.

We focus our attention on the effects of bars on the nuclear activity
and it is very well known that galaxy interactions contribute to the AGN power enhancement
\citep{alonso07}. So, for the purpose of having unbiased results with respect to the bar effects in AGN we 
derived a sample of AGN hosted by relatively isolated barred galaxies.  
In order to construct a sample with suitable isolation criterion for barred 
galaxies we require that any neighboring galaxy within a region of 500 kpc $h^{-1}$ 
in projected separation and $\Delta V>$ 1000 km $s^{-1}$ must be fainter than the barred AGN galaxy.
With this restriction, the final sample of isolated barred 
AGN is composed by 1530 galaxies.

\subsection{Control sample}

\begin{figure}
\includegraphics[width=100mm,height=130mm ]{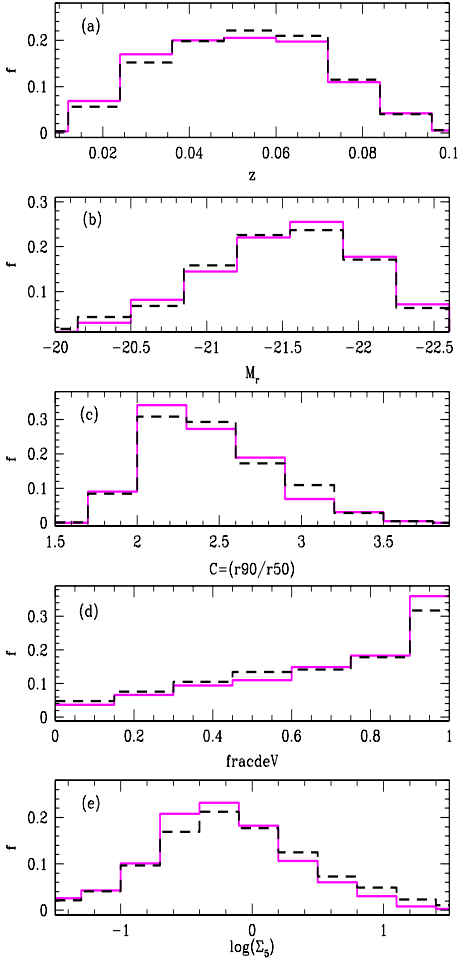}
\caption{Distribution of redshift, luminosities, concentration index, bulge size indicator 
and local density parameter, 
$z$, $M_r$, $C$, $fracdeV$ and $log(\Sigma_5)$ 
($a$, $b$, $c$, $d$ and $e$ panels, respectively), for barred AGN galaxies 
(solid lines) and AGN galaxies without bars in the control sample (dashed lines).}
\label{cont}
\end{figure}

In order to provide a suitable quantification of the effects of bars on 
active galactic nuclei, we constructed a control
sample of unbarred AGN hosts to confront by comparison with the barred host results.
First,  from the sample of 4638 unbarred AGN spiral face-on galaxies (with  $b/a>0.4$ and $g-$mag$<$16.5) 
we selected isolated  objects, using a suitable isolation criterion defined in the previous section for the 
barred AGN sample.
Then, we defined a sample of control galaxies using a Monte Carlo algorithm that selects
galaxies in the unbarred isolated sample, with similar redshift, $r-$band magnitude and concentration index\footnote{$C=r90/r50$ is the ratio of Petrosian 90 \%- 50\% 
r-band light radii}, $C$, distributions of the barred AGN sample 
(see panels \it{a},  \it{b} and  \it{c} in Fig.\ref{cont}).
We restricted the control sample of unbarred AGN hosts to match 
the concentration parameter $C$ of the AGN barred sample in order to have a similar bulge to disk ratio in both samples.
So, the possible differences in the results are associated with the presence of bars and 
not with the difference in the global morphology.

 \cite{coel11} found that bulges in barred galaxies have a lower mass than in unbarred galaxies with similar total stellar mass distribution. Since this tendency could be affect our future analysis, we have considered galaxies with similar bulge prominence in both samples and we have selected the control galaxies with a similar distribution of the $fracdeV$ parameters of the barred AGN host sample 
(panel $d$ in Fig.\ref{cont}).
We notice that the SDSS $fracdeV$ parameter is a good indicator of bulge sizes in galaxies with disk morphology 
\citep{kuehn05,ber10,master10,skibba12} and that
\cite{master10} conclude that this result is independent of the presence of bars. 

We have also selected a control sample with similar distribution of local density 
environment to that of barred galaxies. 
With this aim, for both isolated samples (barred and unbarred) 
we defined a projected local density parameter, $\Sigma_5$.
This parameter is calculated through the projected distance $d$ 
to the $5^{th}$ nearest neighbor, $\Sigma_5 = 5/(\pi d^2)$.
Neighbors have been chosen to have luminosities above a certain  threshold
and with a radial velocity difference less than 1000 km $s^{-1}$.
We also imposed the condition $M_r < -20.5$ to select   in SDSS \citep{balo04}.
Panel ($e$) in Fig.\ref{cont} shows the $log(\Sigma_5)$ distribution for both samples.

The procedure followed to construct this control catalog
assures that it will have the same selection effects as the barred AGN catalog, and consequently, 
it can be used to  estimate the actual difference between AGN galaxies with and without bars,  unveiling 
the effect of the bars on the nuclear activity driven by the central black hole feeding.

\begin{figure*}
\includegraphics[width=165mm,height=130mm ]{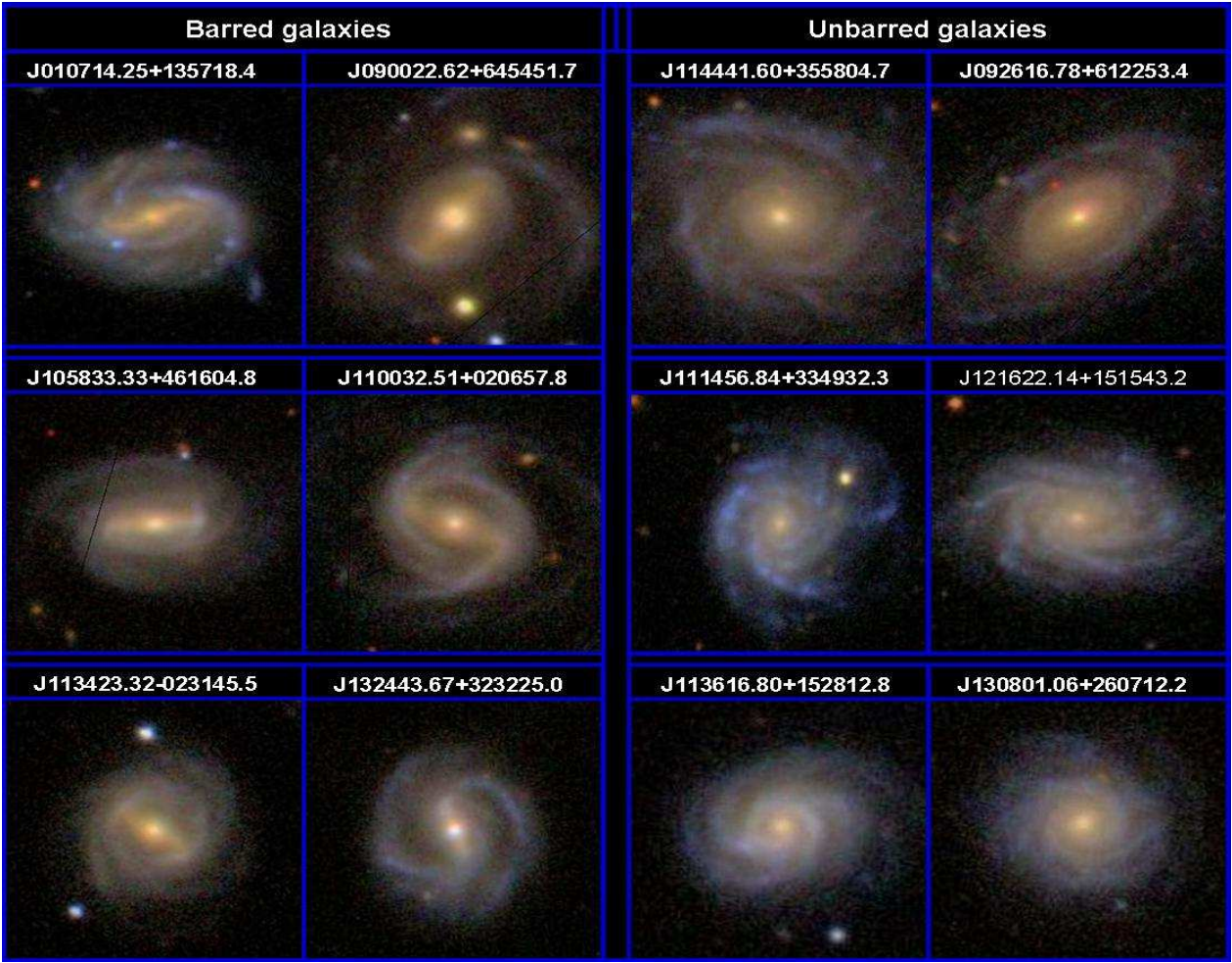}
\caption{Images of typical examples of barred and unbarred AGN galaxies in our samples.
The image sizes are 35 x 35 kpc$^2$.}
\label{ej}
\end{figure*}

Fig.\ref{ej} shows images of typical examples of barred and unbarred AGN galaxies 
(left and right panels, respectively) selected for this work.

\section{AGN host properties}

Different studies have shown that the bars can induce several processes which modify many 
properties of the galaxies (e.g. \cite{sell93,comb93,zari94}) and also can have associated 
different star formation history.
The bars can trigger nuclear starbursts \citep{gado01}, modify the galactic structure 
\citep{atha83,buta96}, change the chemical composition and dynamics 
of gas \citep{martin94,comb93}, etc. 
With the purpose of analyzing the impact of bars on the AGN host properties, 
in this section we explored the effect of bars on the stellar population, color index and 
local density environment of the AGN host galaxies.

\begin{figure}
\includegraphics[width=80mm,height=90mm ]{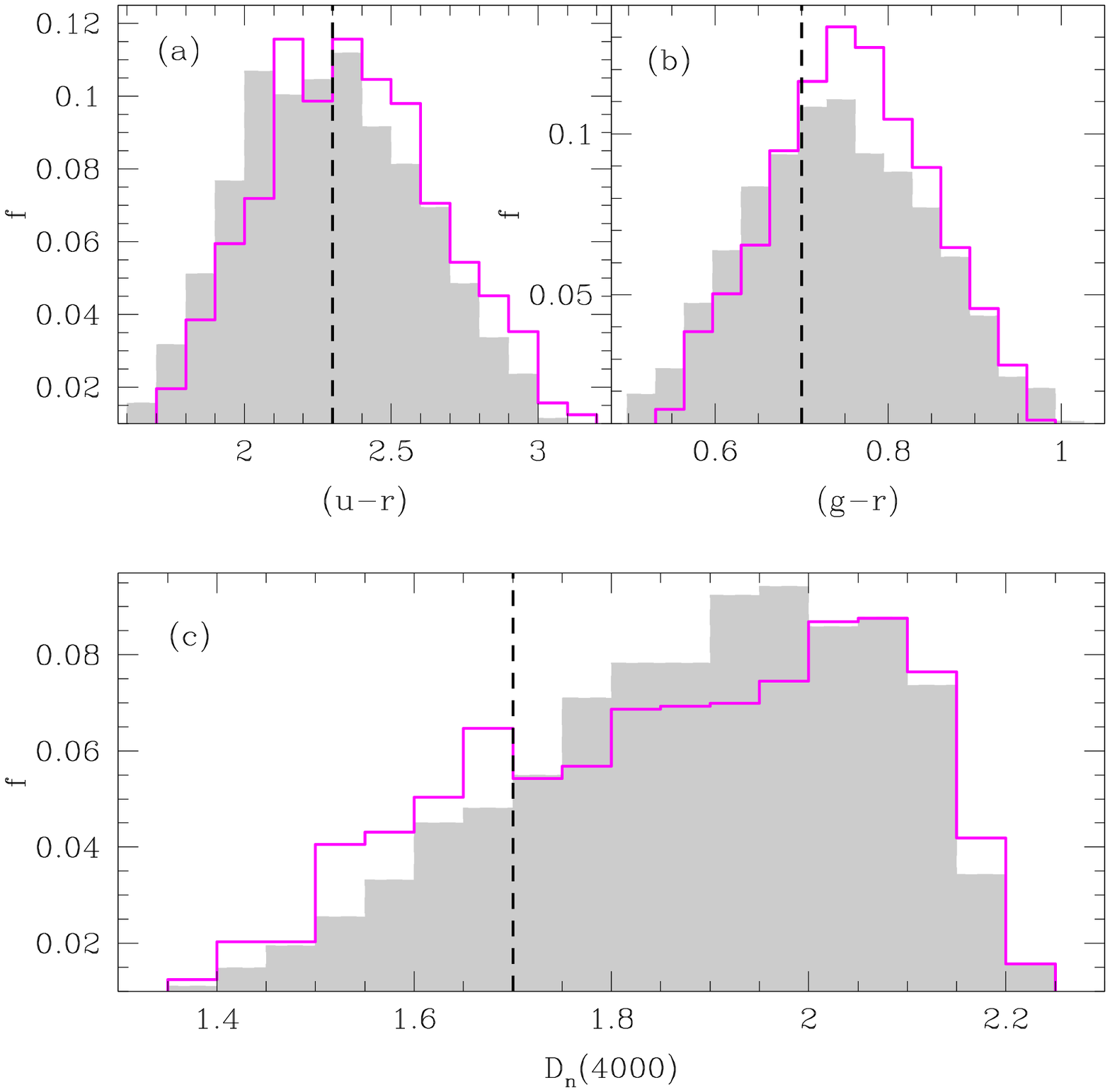}
\caption{Distribution of colors, $(u-r)$ and $(g-r)$, (panels $(a)$ and $(b)$, respectively) 
and stellar age indicator, $D_n(4000)$, ($c$ panel) 
for barred AGN galaxies (solid lines) and AGN galaxies without bars (shaded histogram).}
\label{ColDn}
\end{figure}

In Fig.~\ref{ColDn} (upper panels) we show the $(u-r)$ and $(g-r)$  color distributions of 
barred and unbarred AGN galaxies. 
It can be seen that there is a clear excess of barred AGN hosts with redder colors. 
In the same direction, recent results of \cite{master10} found that passive red spiral galaxies had a high
fraction of bars, within a sample obtained from the Galaxy Zoo catalog.
Furthermore, \cite{oh12} also found that a significant number of barred
galaxies are redder than typical late-type galaxies, and  
the majority of unbarred spiral galaxies are
located in the blue peak.
Our results on barred AGN galaxies are consistent with those of 
barred galaxies without nuclear activity found in previous works.

To complement this analysis, we also use the spectral index $D_n(4000)$ 
as an indicator of the age of stellar populations. 
This spectral discontinuity occurring at 4000 $\AA $ \citep{kauff03} arises by an
accumulation of a large number of spectral lines in a narrow region of the spectrum, an effect 
that is important in the spectra of old stars.
We have adopted \cite{balo99} definition of  $D_n(4000)$ as the ratio of the average 
flux densities in the narrow continuum bands (3850-3950 $\r{A}$ and 4000-4100 $\r{A}$). 
 In the lower panel of Fig.~\ref{ColDn} we show the distribution of $D_n(4000)$ values for barred AGN and galaxies without bars in the control sample.
We find an excess of barred objects exhibiting 
lower values of $D_n(4000)$, showing that the hosts of barred AGN galaxies have an important 
young stellar population. The differences in the colour and $D_n(4000)$ distributions of barred 
and unbarred AGN (Fig. 3) were confirmed using the Kolmogorov-Smirnov test with a significance 
higher than 99,95\%.

According to these results, we define a population of red and young hosts by 
restricting to values of colors and $D_n(4000)$ parameters for the barred AGN sample. The 
thresholds are indicated by the dashed vertical line in Fig.\ref{ColDn}.
In Table 2 we quantify this excess of red colors and young stellar population of barred AGN 
galaxies with respect to the control sample.

\begin{table}
\center
\caption{Percentages of red colors and young stellar population for barred AGN galaxies,
 with respect to unbarred AGN in the control sample.}
\begin{tabular}{|c c c c| }
\hline
Restrictions & $(u-r)>2.3$ &  $(g-r)>0.72$ & $D_n(4000)<1.7$  \\
\hline
\hline
$\%$ of barred    &   58.1$\%$ & 63.6$\%$ & 30.1$\%$ \\
$\%$ of unbarred  &   49.3$\%$ & 54.2$\%$ & 22.3$\%$ \\
\hline
\end{tabular}
{\small}
\end{table}

\begin{figure}
\includegraphics[width=95mm,height=93mm ]{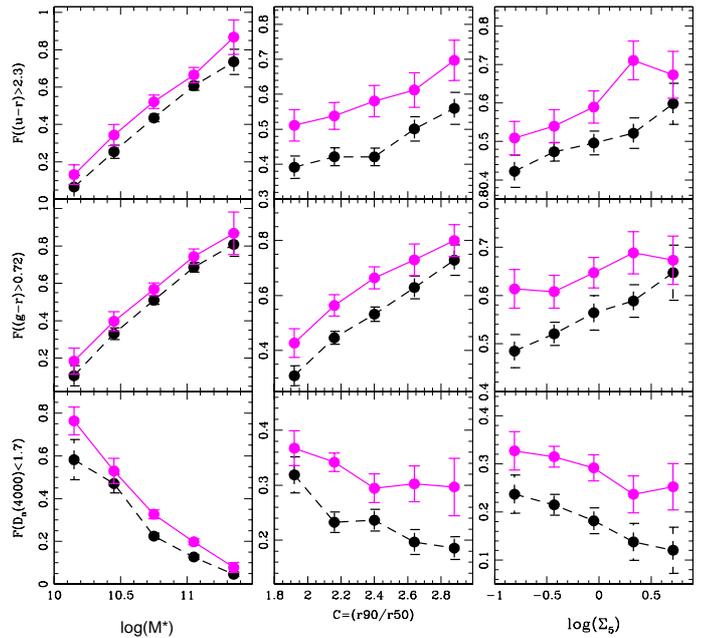}
\caption{Fraction of galaxies with $(u-r)>2.3$ (upper panel), 
$(g-r)>0.73$ (middle panel) and $(D_n(4000))<1.7$ (lower panel), as a function of the stellar mass, 
concentration index and local environment parameter, $log(\Sigma_5)$, 
(left, middle and right panels, respectively),  
for barred and unbarred AGN galaxies (solid and dashed lines, respectively).}
\label{DnColdc}
\end{figure}

In order to understand the behavior of the colors and stellar populations of barred AGN 
with respect to AGN galaxies without bars, we have analysed the relative fractions of red and young stellar populations.
In Fig.~\ref{DnColdc} we show the fraction of $u-r > 2.3$ (upper panel), $g-r > 0.72$ 
(middle panel) and $D_n(4000)< 1.7$ (lower panel) as a function of stellar mass, $log(M^*)$, 
concentration index, $C$, and local environment parameter, $log(\Sigma_5)$, 
for barred AGN galaxies and for the control sample (solid and dashed lines, respectively).
Errors have been estimated by applying the bootstrap resampling technique in 
all figures \citep{barrow84} the error-bar length correspond to 1$\sigma$.
As can be seen in the left panels, the fraction of red AGN galaxies 
increases towards high stellar masses and, on the other hand, the fraction of young stellar population decrease with $log(M^*)$.
It can be seen that barred 
AGN host galaxies have a systematically higher fraction of redder colors and younger stellar population in different bins of stellar mass, suggesting that bars have an important effect in producing a rapid transformation of gas into stars associated to reddening and a young population of stars.

Middle panels of Fig.~\ref{DnColdc} show that the red/young fraction
increase/decrease toward the more concentrated host galaxies, for both barred and unbarred AGN. 
This result is consistent with expectations, since galaxies with higher values of the concentration 
index are related to the bulge type morphology, and lower concentration objects, to spiral galaxies.
It can also be seen that barred AGN galaxies (solid lines) show higher fractions of red and young population, for the whole range of the $C$ parameter compared to their counterpart of unbarred AGN in the control sample.

Regarding to the relation between bars and environment, 
different works find that local density could be important in the formation of bars, 
although results are contradictory.
Numerical simulations (e.g. \cite{walk96,mihos97,bere04}) show that bars are created by 
interactions between galaxies. 
Different observational studies find that the bar galaxy fraction does not depend on the environment 
\citep{vandenB02,mendez10,martinez11}. 
More recently, \cite{lee12} also find that the fraction of weak bars has no correlation with environmental 
parameters, suggesting that there is no direct evidence for environmental stimulation of bar formation.
On the other hand, \cite{elme90} show a correlation between the bar fraction and environment for
 spiral galaxies, finding the highest fraction in pair systems.  
More recently, \cite{skibba12}, from a sample obtained from the Galaxy Zoo 2 project, clearly detected the 
environmental dependence of barred galaxies, such that barred tend to be found in denser environments than their 
unbarred counterparts, with environmental correlations that are statistically significant 
(at a level of $>$ 6$\sigma$) on scales of 150 kpc to a few Mpc. 
The authors argue that the small number statistics of previous studies inhibited 
their detection of a bar-environment correlation.

In order to study the dependence of colors and stellar populations of AGN hosts on the environment, we show in the right panels of  
Fig.~\ref{DnColdc} the red, young population fractions of barred AGN hosts in the control sample as a function of the local galaxy density parameter ($\Sigma_5$). 
It can be seen a weak trend consistent with the expected increase of 
the red fractions with local galaxy density for AGN hosts with and without bars.
Nevertheless, the fraction of low $D_n$ hosts, where an important young stellar population is present, decreases towards higher values of $\Sigma_5$.
Interestingly, in the whole range of $log(\Sigma_5)$, the red and young population fractions of galaxies in barred AGN 
galaxies (solid lines) exceeds to that of unbarred AGN in the control sample (dashed lines).
According to the boostrap error bars this signal is statistically meaningful at more than
$1\sigma$-level in left panels, and $2\sigma$-level in middle and right panels.

From these analysis we conclude that the dependence of colors and stellar populations on environment suggests that bar perturbations are a suitable physical mechanism for a fast 
star formation activity and a stirring of dust in the central regions of these AGN host galaxies.

\section{Black hole activity}

As a tracer of the AGN activity, we focus here on the dust corrected luminosity of 
the [OIII]$\lambda$5007 line, $Lum[OIII]$.
This estimator is widely used by several authors  \citep{mul94, kauff03, heck04, heck05, 
brinch04} and some catalogs 
of [OIII]$\lambda$5007 luminosities have been published \citep{whitt92, xu99}
The main reason is that the [OIII] line is
particularly appealing because it is one of the strongest narrow emission 
lines in optically obscured AGN and with very low contamination by contributions of 
star formation in the host galaxy. 
AGN hosts are typically massive galaxies
so their metallicities are expected to be high due to the mass-metallicity relation \citep{kauff03, 
tremonti04}. For our sample, the majority of the  galaxies have $M^* >10^{10}$ $M_{\sun}$ so that higher 
metallicities are expected. \cite{kauff03} shows by inspection to the BPT diagrams that high 
metallicity hosts have a low contamination in the $Lum[OIII]$ by star formation. 
Furthermore, \cite{heck04}, compute the average contribution of the AGN to the $Lum[OIII]$ for a 
sample of composite galaxies (AGN above the  \cite{kauff03} criterium but below the
\cite{kew01} 2001 line in the BPT diagram) and the AGN-dominated galaxies (above  the \cite{kew01} 
2001 demarcation in the BPT diagram) taken from SDSS. The results of this work indicate that
the $Lum[OIII]$ emission from composite galaxies comes, in a range of 50\% to 90\% from 
AGN activity, while for AGN-dominated galaxies, more than 90\% of the $Lum[OIII]$ comes
from AGN emission.
In addition, \cite{heck05} found a correlation between the hard X-ray 
emission of AGN and the OIII luminosity concluding that, at low redshifts, the selection 
by optical emission lines will recover most AGN selected by hard
X-ray emission.

\begin{figure}
\includegraphics[width=70mm,height=60mm ]{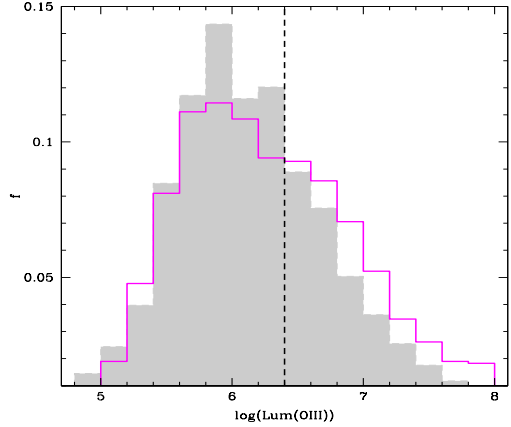}
\caption{Distribution of $log(Lum[OIII])$ for barred AGN galaxies (solid line) and unbarred AGN (full surfaces).
}
\label{hLO}
\end{figure}

The influence of the bars on nuclear activity can be seen in Fig.\ref{hLO},
 in the histogram of $log(Lum[OIII])$ for barred AGN (solid line), 
and galaxies in the control sample (full surface).
It can be appreciated that barred AGN galaxies show an excess of high $Lum[OIII]$ values 
with respect to the control sample.
The Kolmogorov-Smirnov statistics allow us to quantify the difference between these distributions with a significance of 99,99\%.
Here, we consider the luminosity $Lum[OIII]= 10^{6.4} L_{\odot}$ as a limit 
between weak and powerful AGN, by selecting values that represent the excess of 
nuclear activity for the barred AGN sample 
($43.4\%$ for barred AGN galaxies versus $31.5\%$ for the control sample).
This threshold is represented by the dashed vertical 
line in Fig.\ref{hLO}.

\begin{figure*}
\includegraphics[width=160mm,height=200mm ]{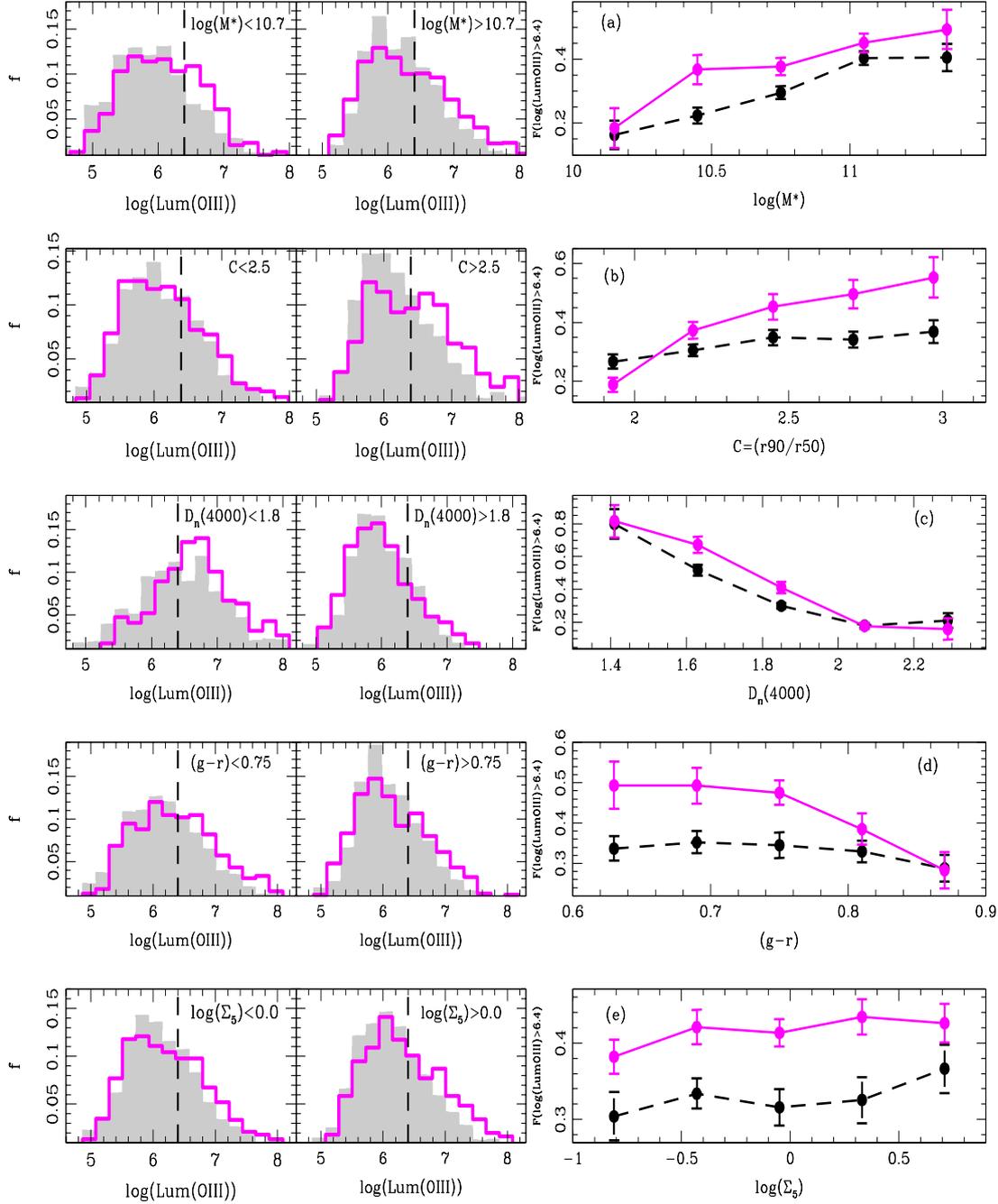}
\caption{Distributions of $log(Lum[OIII])$ for barred and unbarred AGN galaxies (solid
lines and full surfaces, respectively), in different ranges of $log(M^*)$, $C$, $D_n(4000)$, $(g-r)$ and 
$log(\Sigma_5)$.
Right panels show fraction of galaxies with $log(Lum[OIII])>6.4$ as a function of stellar masses ($a$),
 concentration parameter ($b$), stellar age population ($c$),  $(g-r)$ color ($d$) and
local density environment ($d$) for AGN barred galaxies and unbarred AGN (solid and dashed lines, respectively).
}
\label{LOAll}
\end{figure*}

Left and middle panels in Fig.~\ref{LOAll}, show the nuclear activity distributions, $log(Lum[OIII])$, 
of barred AGN galaxies in comparison with the unbarred AGN counterparts. 
The analysis was performed for
the different ranges of stellar masses, $log(M*)$, concentration parameter, $C$, stellar age indicator, 
$D_n(4000)$,  $(g-r)$ color and local environment, $log(\Sigma_5)$.
We find that in more massive galaxies, with older stellar populations and 
red colors, the $Lum[OIII]$ distributions of barred AGN are similar to those of AGN without bars 
in the control sample.
Interestingly, in less massive, younger stellar population and bluer   
host galaxies ($\approx$ $M^*< 10^{10.7} M_{\odot}$, $D_n(4000) < 1.8$ and $(g-r)<0.75$), the nuclear 
activity distributions of barred and unbarred AGN exhibit significant differences, with barred objects 
having a larger fraction of powerful active nuclei galaxies.
This behavior can be interpreted in terms of the larger ability of bars to fuel the central
black holes in galaxies with a larger gas fraction expected in these hosts.
Moreover, we explore the nuclear activity distributions
of barred and unbarred AGN with different concentration parameter.
The critical concentration index value of $C=2.5$  is
 adopted to segregate concentrated, bulge-like ($C >2.5$) galaxies from more extended, 
disc-like ($C<2.5$) objects.  
Bearing this in mind, we find that in AGN hosts with higher values of $C<2.5$, indicating bulge type
morphology, barred galaxies show a higher fraction of powerful AGN 
in comparison with those of unbarred AGN in the control sample.

We have also selected two different local environment ranges, and it can be appreciated that there are no  differences in the nuclear activity distributions.
In addition, in Table 3 we computed the percentage of barred and unbarred AGN galaxies with $log(Lum[OIII])>6.4$, 
in different ranges of stellar masses, $C$ index, age populations, $(g-r)$ colors and $\Sigma_5$ parameter.  
These percentages reflect the results of $Lum[OIII]$ distributions (Fig. 6) confirming that
most powerful AGN tend to inhabit in blue and young barred host galaxies but with 
a higher value in the concentration index.

To reinforce the results obtained with distributions, in the   
right panels of Fig.~\ref{LOAll} we present the fraction of galaxies 
with $log(Lum[OIII])>6.4$ as a function of galaxy properties. 
Panel ($a$) shows the fraction of galaxies with high $Lum[OIII]$ values, as a function of stellar 
mass content of the corresponding host galaxies.
This relation has been calculated for barred AGN and for AGN galaxies in the 
control sample without bars.
From the results shown in Fig.\ref{LOAll}a, it can be clearly
appreciated that, in general, most massive hosts show the higher 
fraction of AGN with $Lum[OIII] > 10^{6.4} L_{\odot}$.
We can also see that barred AGN objects systematically show a higher fraction of powerful 
AGN galaxies irrespective of the stellar mass content, 
indicating an enhancement of the  black hole activity 
or barred AGN with respect to AGN residing in unbarred hosts.
In the similar way, \cite{oh12} found that the AGN strength is enhanced by the presence of a bar and linearly correlates with stellar mass. 

In panel $b$ we show the fraction of powerful AGN galaxies as a function of concentration 
index, $C$.
In this plot we find that the fraction of barred AGN galaxies with 
$log(Lum[OIII])>6.4$ increases toward 
higher $C$ values, while the fraction of powerful unbarred AGN appear 
to remain almost constant with the morphological parameter.
In addition, it can be appreciated that barred AGN galaxies show, on average, a higher 
fraction of powerful AGN, with respect to the control sample, but with a decrease of powerfully
barred AGN for the lowest values of the $C$ parameter where both barred and unbarred AGN have 
similar fraction of $Lum[OIII]$.

Fig.~\ref{LOAll}c shows the fraction of barred AGN galaxies and of AGN in the control sample, 
with strong $Lum[OIII] > 10^{6.4} L_{\odot}$, as a function of the 
stellar age parameter ($D_n(4000)$). 
From this figure we can see that the trend is consistent with the 
increase of the powerful AGN galaxy fraction with younger stellar population objects, for both, 
barred and unbarred AGN galaxies.
Also, we can appreciate that the fraction of barred AGN galaxies with 
$Lum[OIII] > 10^{6.4} L_{\odot}$ (solid line) is, for younger galaxies, higher than 
that of AGN hosts without bars (dashed lines).
A similar trend is observed with respect to galaxy color $(g-r)$ (see Fig.\ref{LOAll}d) where 
the fraction of barred AGN galaxies with higher nuclear activity decreases toward redder colors
while the fraction of powerful unbarred AGN remains constant for the whole
($g-r$) range.

With the aim of understanding the behavior of the nuclear activity in AGN galaxies with and 
without bars as a function of local density environments, we also measured the powerful AGN
fraction with respect to the $\Sigma_5$ parameter.
The results are shown in Fig.~\ref{LOAll}e where can be observed that 
the fraction of powerful AGN increases slightly toward denser regions. Moreover, 
it is important to notice the significant excess of barred AGN with high $Lum[OIII]$ values with 
respect to the unbarred AGN, independently of the local environment density.

\begin{table}
\center
\caption{Percentages of barred and unbarred AGN galaxies with $log(Lum[OIII])>6.4$, 
in different ranges of stellar masses, $C$, $D_n(4000)$, $(g-r)$ and $log(\Sigma_5)$. 
}
\begin{tabular}{|c c c| }
\hline
Ranges & $\%$ of barred AGN & $\%$ of unbarred AGN \\
\hline
\hline
$log(M^*)>$10.7         & 43.5$\%$  &     36.6$\%$        \\
$log(M^*)<$10.7         & 35.5$\%$  &     22.9$\%$        \\
\hline
\hline
$C<$2.5               & 35.1$\%$  &     31.5$\%$        \\
$C>$2.5               & 51.8$\%$  &     34.4$\%$        \\
\hline
\hline
$D_n(4000)>$1.8         & 24.8$\%$  &     22.5$\%$        \\
$D_n(4000)<$1.8         & 66.3$\%$  &     51.1$\%$        \\
\hline
\hline
$(g-r)>$0.75         & 36.3$\%$  &     29.1$\%$        \\
$(g-r)<$0.75         & 42.3$\%$  &     35.9$\%$        \\
\hline
\hline
$log(\Sigma_5)>$0.0         & 44.3$\%$  &     32.6$\%$        \\
$log(\Sigma_5)<$0.0         & 40.1$\%$  &     32.8$\%$        \\
\hline
\hline
\end{tabular}
{\small}
\end{table}

\subsection{Black Hole Mass and Accretion Rate}

Moreover, it is interesting to further investigate the strength of AGN.
For this aim, we estimated black holes masses, $M_{BH}$, for barred and unbarred AGN galaxies, 
by using the observed correlation between $M_{BH}$ and the
bulge velocity dispersion $\sigma_*$ (Tremaine et al. 2002).

\begin{equation}
logM_{BH} = \alpha + \beta log(\sigma_*/ 200)
\end{equation}

The $M_{BH} - \sigma_*$ relation is different for barred and unbarred galaxies, because the central velocity dispersion is enhanced by the stellar motion along a bar \citep{gra08}.
We adopted ($\alpha$, $\beta$) = (7.67 $\pm$ 0.115, 4.08 $\pm$ 0.751) for barred AGN galaxies 
and ($\alpha$, $\beta$) = (8.19 $\pm$ 0.087, 4.21 $\pm$ 0.446) for unbarred AGN.
\citep{gulte09}. 
We restricted this analysis to AGN galaxies with $\sigma_*$ $>$ 70 km
$s^{-1}$, because the instrumental resolution
of SDSS spectra is $\sigma_*$ $\approx$ 60 to 70 km $s^{-1}$.

\begin{figure}
\includegraphics[width=90mm,height=80mm ]{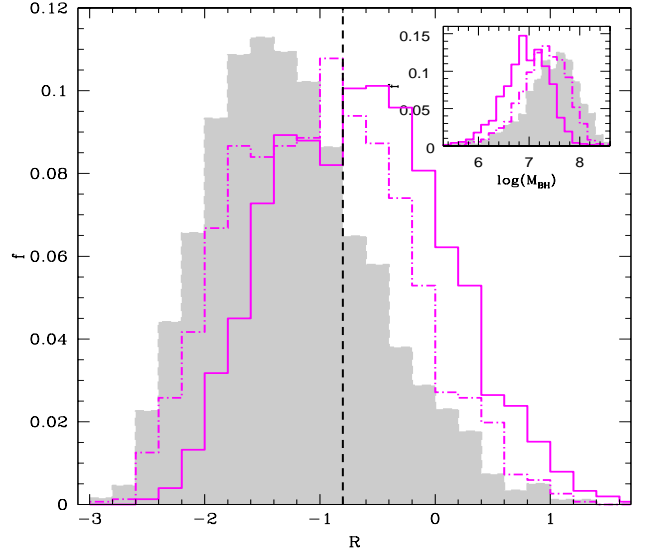}
\caption{Distribution of $\cal R$ for barred AGN galaxies (solid line) and 
unbarred AGN (full surfaces).
The small box correspond to the distribution of $log(M_{BH})$.
Dot-dashed lines represent the $log(M_{BH})$ and $\cal R$ distributions for barred AGN galaxies, 
 using the same parameters $\alpha$ and $\beta$, of the unbarred control sample. 
}
\label{hR}
\end{figure}

Furthermore, the ratio of [OIII] luminosity to black hole mass 
($\cal R$=log($Lum[OIII]$/$M_{BH}$)) provides a useful measure of the accretion rate onto a black hole 
\citep{heck04}.
The big panel in Fig.\ref{hR} shows the $\cal R$ distributions for barred (solid line) 
and unbarred AGN (full surface).
It can be appreciated that barred AGN galaxies show an excess of high accretion rate values 
with respect to the control sample.
With the purpose of checking the effect of bar in the active galactic nuclei in more detail, 
we also calculated the accretion rate for barred AGN using the same parameters, $\alpha$ and $\beta$, 
that we used to obtained $log(M_{BH})$ and $\cal R$ of the unbarred AGN in the control sample. 
It is represented by dot-dashed lines in Fig.\ref{hR}. Although, in this last case, the signal is less significant,
also, barred AGN galaxies show an excess of accretion rate compared to the AGN without bars.

Besides, in the small box in Fig.\ref{hR}, we show the distributions of
black holes masses, $M_{BH}$, for barred and unbarred AGN galaxies (solid line and full surface, respectively).
It can be appreciated in this figure that there is a trend for
$M_{BH}$ to be systematically larger for unbarred AGN.
As can be observed, the accretion rate and black hole mass distributions for barred and unbarred AGN calculated 
with the same parameters show a difference of lower statistical significance (K-S test with 99,95\% and 98,8\% confidence, 
respectively) than in the case when barred AGN black hole masses are calculated with the specific $\alpha$, $\beta$ parameters 
(K-S test with 100\% confidence for both  $\cal R$ and $M_{BH}$ distributions).

These results show that the excess of the accretion rate within a black holes 
in barred AGN galaxies with respect to unbarred AGN is a real effect, 
indicating that  bars play an important role in the gas infall toward the central 
regions in active nuclei galaxies.

\section{Summary and Conclusions}

We have performed a statistical analysis of both, host characteristics and nuclear activity, 
of AGN galaxies with and without bars. 
Our study is based on the sample of AGN derived from the SDSS-DR7 release, using the publicly available emission-line fluxes.
We have complemented the SDSS-DR7 data with the addition of a naked-eye detection of 
barred features of images of 6772 face-on AGN spiral galaxies brighter than $g-$mag$<$16.5. 
With the aim of providing an appropriate quantification of the effects of bars, 
we constructed a suitable control sample of unbarred AGN galaxies, 
with the same redshift, $r-$band magnitude, concentration index, 
bulge size parameter and local environment distributions. 

We can summarize the main results in the following conclusions.

(i)  We found 1927 barred AGN, with respect to the full sample of 
6772 spiral face-on AGN host galaxies, which represents a fraction of 28.5$\%$.
This result agrees well with the bar 
fraction found by visual inspection of optical galaxy samples in previous works 
(e.g. \cite{nilson73,marinova09,master10,oh12}) but is lower than that found using imagen descomposition 
technique \citep{gado09} and near-infrared analisys images \citep{menendez07}.
We also cross-correlated our sample with barred AGN galaxies taken from the catalog of Nair \& Abraham (2010), and we found that the visual classification of both catalogs is in very good agreement.

(ii) We examined the color distribution of barred and unbarred AGN galaxies, 
and we found that there is an important excess of barred AGN hosts with red colors. 
The distribution of stellar population of
barred AGN sample shows an excess of young galaxies, with respect to the control sample.
Redder colors and younger stellar populations found in barred AGN galaxies 
suggest that bar perturbations are an important effect in modifying galaxy properties in the hosts of AGN, producing a significant star formation activity and a stirring of dust in the central regions.

(iii) We also studied the fraction of galaxies with red colors and young stellar 
population as a function of  $log(M^*)$, concentration index and local density parameter, in barred AGN galaxies and in the control sample.
We found that the number of red/young host galaxies increases/decreases with higher values of stellar mass, concentration parameter and local environment.
It can be also seen that barred AGN galaxies show a higher fractions, with respect to their counterpart of unbarred AGN in the control sample.

(iv) We found that barred AGN galaxies show an excess of higher nuclear activity, with respect to AGN without bars in the control sample.
In addition, we find that this tendency is more significant in less massive, 
more concentration light, younger stellar population and bluer AGN host galaxies.

(v) We also explored the fraction of powerful AGN (with $Lum[OIII]>10^{6.4} L_{\odot}$) 
as a function of the host galaxy properties.
We find that the fraction of AGN galaxies with $Lum[OIII]>10^{6.4} L_{\odot}$ increase 
toward more massive host, higher C values, bluer colors, younger stellar population objects 
and denser environments.
We also found that barred AGN objects systematically show a higher fraction of powerful 
AGN with respect to the control sample, in bins of different galaxy properties.

We also analyzed the accretion rate onto a black hole for barred and unbarred AGN galaxies. From this study we concluded that barred AGN have an excess of objects with high accretion rate values with respect to the control sample. 
This result suggests that gas rich galaxies are more efficient in accretion material toward the central region and also imply that bars can help in the fueling of material onto central black hole.

\begin{acknowledgements}
      This work was partially supported by the Consejo Nacional de Investigaciones
Cient\'{\i}ficas y T\'ecnicas and the Secretar\'{\i}a de Ciencia y T\'ecnica 
de la Universidad Nacional de San Juan.

Funding for the SDSS has been provided by the Alfred P. Sloan
Foundation, the Participating Institutions, the National Science Foundation,
the U.S. Department of Energy, the National Aeronautics and Space
Administration, the Japanese Monbukagakusho, the Max Planck Society, and the
Higher Education Funding Council for England. The SDSS Web Site is
http://www.sdss.org/.

The SDSS is managed by the Astrophysical Research Consortium for the
Participating Institutions. The participating institutions are the American
Museum of Natural History, Astrophysical Institute Potsdam, University of
Basel, University of Cambridge, Case Western Reserve University,
University of
Chicago, Drexel University, Fermilab, the Institute for Advanced Study, the
Japan Participation Group, Johns Hopkins University, the Joint Institute for
Nuclear Astrophysics, the Kavli Institute for Particle Astrophysics and
Cosmology, the Korean Scientist Group, the Chinese Academy of Sciences
(LAMOST), Los Alamos National Laboratory, the Max-Planck-Institute for
Astronomy (MPIA), the Max-Planck-Institute for Astrophysics (MPA), New Mexico

State University, Ohio State University, University of Pittsburgh, University
of Portsmouth, Princeton University, the United States Naval Observatory, and
the University of Washington.
\end{acknowledgements}

\end{document}